\documentclass{article}

\newcommand{\p}[1]{(\ref{#1})}

\newcommand{\cF}{{\cal F}}

\newcommand{\be}{\begin{equation}}
\newcommand{\ee}{\end{equation}}
\newcommand{\bea}{\begin{eqnarray}}
\newcommand{\eea}{\end{eqnarray}}

\newcommand{\ba}{\begin{array}} \newcommand{\ea}{\end{array}}

\def\im{{\rm i}}

\newcommand{\nn}{\nonumber}
\def\theequation{\arabic{section}.\arabic{equation}}

\usepackage{amscd,amsmath,amssymb}

\begin{document}
\thispagestyle{empty}
\vspace{2cm}
\begin{flushright}
\end{flushright}\vspace{2cm}
\begin{center}
{\Large\bf Supersymmetrizing the Pasti-Sorokin-Tonin action}
\end{center}
\vspace{1cm}

\begin{center}
{\large\bf  N.~Kozyrev${}^a$}
\end{center}

\begin{center}
${}^a$ {\it
Bogoliubov  Laboratory of Theoretical Physics, JINR,
141980 Dubna, Russia} \vspace{0.2cm}
\end{center}
\vspace{2cm}

\begin{abstract}\noindent
In this paper the harmonic superspace action of the tensor multiplet of $N=(1,0)$, $d=6$ supersymmetry is constructed which in the bosonic limit reduces to the known Pasti-Sorokin-Tonin action for the self-dual tensor field. The action involves, besides the potential containing the dynamical fields, also an auxiliary tensor multiplet and a set of analytic superfields with gauge PST scalar among them. For each of gauge symmetries of the PST action, a superfield analog is found. The equations of motion are calculated and it is shown in the bosonic limit that no extra degrees of freedom appear.
\end{abstract}

\setcounter{page}{1}
\setcounter{equation}{0}


\section{Introduction}
Tensor multiplets of six-dimensional supersymmetry are known to play an important role in supersymmetric and string theories. First, the fields of $N=(2,0)$ multiplet effectively describe excitations of $M5$ brane, with the action given by Born-Infeld-type generalization of the free action of this multiplet \cite{M5brane}. It is anticipated that the nonabelian generalizations of this action describe stacks of $M5$ branes. In the lowest energy limit, when all Born-Infeld nonlinearities could be neglected, stacks of $M5$ branes are supposed to be described by $N=(2,0)$, $d=6$ superconformal field theory, maximally possible superconformal theory, which is thus also related to the tensor multiplet. Conclusion about this relation can also be drawn from study of representations of superconformal algebras \cite{nahm}. Attempts to find the nonabelian generalization of the tensor multiplet, its action, tensor hierarchies, typically involve studies of on-shell multiplets or even components \footnote{See, for example, \cite{saemann, samtleben}}, which makes this rather nontrivial task even more complicated. This makes desirable to find off-shell framework to develop such theories.

Even if one limits oneself to just $N=(1,0)$, $d=6$ supersymmetry, the standard superfield approach to the tensor multiplet allows to define it only on-shell and is not useful in construction of the superfield action. Therefore, one has to employ more elaborate approaches, such as harmonic superspace \cite{HS,d6HS}. The free action that was constructed in this superspace involves not just one but two different multiplets, acting as Lagrangian multipliers to each other \cite{sokatchev}. It was proposed to use the fields of supergravity multiplet as auxiliaries to identify these two \cite{sokatchev}. Requirement of using two multiplets is not surprising if one recalls that the tensor multiplet involves the two-form gauge field, which field strength is self-dual on-shell, and such fields do not exist off-shell on their own \cite{sokatchev, bergshoeff}.

To construct the proper action for the tensor multiplet, one can try to find supersymmetric generalizations of known bosonic actions that produce the self-duality equation as the equation of motion. The long study of self-dual fields resulted in a number of approaches, such as noncovariant actions \cite{henneaux,perry}, introduction of infinite tower of Lagrange multipliers \cite{inffields1,inffields2,inffields3}, Pasti-Sorokin-Tonin approach \cite{PST} that involves one auxiliary scalar, and Sen's approach which uses the self-dual 3-form in $6$ dimensions \cite{sen1,sen2}. The PST approach seems to be favored among these for being explicitly Lorenz- and gauge-covariant, not introducing any extra physical degrees of freedom to the theory, though being highly nonlinear. This nonlinearity prevented construction of the superfield PST-type action for the tensor multiplet so far. However, recently the reformulation of the PST approach was introduced \cite{mkrtchyan}, which involves a scalar and an auxiliary 2-form field. The action in this approach reduces to the PST action upon removal of the auxiliary field by its equation of motion and is polynomial, which greatly simplifies its supersymmetrization.

The purpose of this paper is to construct the $N=(1,0)$ harmonic superspace action for the tensor multiplet, which reduces to the PST action in the bosonic limit upon removal of auxiliary fields. There are two reasons to limit this work to $N=(1,0)$ supersymmetry. First, harmonic superspace technique is not powerful enough to construct the action with manifest $N=(2,0)$ supersymmetry. One can achieve it only in hidden way by proper coupling the $N=(1,0)$ theory to the hypermultiplet, which is out of scope of this paper. Second, it is likely that for the nonabelian theory only $N=(1,0)$ part of supersymmetry could be realized, much like the case of BLG and ABJM models, with the latter possessing only $N=6$, $d=3$ supersymmetry instead of $N=8$ \cite{saemann}.

\setcounter{equation}0
\section{The Pasti-Sorokin-Tonin action}
The action that produces self-duality equations of motion for tensor fields in even-dimensional spacetime was proposed by Pasti, Sorokin and Tonin in \cite{PST}. Its six-dimensional version reads
\be\label{PSTact}
S_{PST} = \int d^6 x \Big(  \frac{1}{6}F_{MNP}F^{MNP} -\frac{1}{2 \partial^K z \partial_K z} \cF_{ABC}\cF^{ABD}\partial^C z \partial_D z    \Big), \;\; \cF_{ABC} = F_{ABC} - \frac{1}{6}\epsilon_{ABCMNP}F^{MNP}.
\ee
Here $F_{ABC} = \partial_A B_{BC} - \partial_B B_{AC} + \partial_C B_{AB}$ is the field strength of the 2-form potential $B_{AB}$.

Unlike other proposals to solve the problem of contruction of action of self-dual fields, the PST action combines a set of useful properties:
\begin{itemize}
\item It is manifestly Lorentz-covariant, unlike actions proposed in \cite{henneaux,perry};
\item It contains only a finite number of fields, unlike actions with an infinite set of Lagrange multipliers \cite{inffields1,inffields2,inffields3};
\item It does not contain any extra physical degrees of freedom, and one does not need to show that they decouple, unlike Sen's approach \cite{sen1,sen2}.
\end{itemize}
The third property is not spoiled by the field $z$, as it is a purely gauge degree of freedom. Indeed, the action \p{PSTact} possesses a set of gauge symmetries
\bea\label{PSTtrans}
&& 1) \;\; \delta_f B_{MN}=\partial_M f_N - \partial_N f_M,\;\; \delta_f z=0, \nn \\
&& 2) \;\; \delta_a B_{MN} = \partial_{[M} z \; a_{N]}, \;\; \delta_a z=0, \\
&& 3) \;\; \delta_\lambda z =\lambda, \;\;  \delta_\lambda B_{MN} = \frac{\lambda}{\big(\partial z\big)^2} \cF_{MNP}\partial^{P}z.\nn
\eea
The last transformation \p{PSTtrans} allows to fix the field $z$. By setting $z =x^5$, one can reproduce action proposed in \cite{henneaux,perry}. Equivalence of \p{PSTact} and the action with infinite number of fields was shown already in \cite{PST}.

The supersymmetric component version of \p{PSTact} was already used in studies of nonabelian version of this theory \cite{saemann}. The highly nonlinear nature of this action makes it difficult, however, to find its superfield form. The idea how to circumvent this comes from the work by Mktrchyan \cite{mkrtchyan}, where polynomial version of \p{PSTact} was constructed
\bea\label{PSTMact}
S_{PSTpoly} &=& \int d^6 x \Big(  \frac{1}{6}F_{MNP}F^{MNP} +\frac{1}{6}\big( \cF_{MNP} -3 \partial_{[M}z R_{NP]}  \big) \big( \cF^{MNP} -3 \partial^{[M}z R^{NP]}  \big)  \Big)= \nn \\
&=&  \int d^6 x \Big( \frac{1}{6}F_{MNP}F^{MNP} - \cF_{MNP}\partial^{M}z R^{NP} +\frac{3}{2}\partial_{[M}z R_{NP]}\partial^{[M}z R^{NP]} \Big).
\eea
Here $R_{[MN]}$ is the auxiliary tensor field. Upon its exclusion by algebraic equation of motion, the action reduces to \p{PSTact}. Just as \p{PSTact}, \p{PSTMact} possesses a set of gauge symmetries:
\bea\label{PTSMgaugetr}
&&1)\;\;\delta_f B_{MN}=\partial_M f_N - \partial_N f_M, \;\; \delta_f R_{MN}=0, \;\; \delta_f z =0, \nn \\
&&2) \;\; \delta_a B_{MN} = \partial_{[M} z \; a_{N]}, \;\;  \delta_a R_{MN} = -\partial_{[M} a_{N]}, \;\; \delta_a z =0, \nn \\
&&3)\;\; \delta_b B_{MN} =0, \;\;  \delta_b R_{MN} = \partial_{[M}z \, b_{N]},\;\; \delta_b z=0,   \\
&&4) \;\; \delta_\lambda z =\lambda, \;\; \delta_\lambda B_{MN} = \lambda R_{MN}, \;\;  \delta_\lambda R_{MN} = \frac{3\lambda }{(\partial z)^2}\Big( \partial_{[M}R_{NP]} + \frac{1}{6}\epsilon_{MNPRST} \partial^R R^{ST}  \Big)\partial^P z. \nn
\eea
Let us note that the action above can also be written in spinor notation,\footnote{To relate these two notations, we use $\gamma$-matrices with properties defined in \cite{buchbinder}} which is most natural and convenient when it comes to supersymmetry. In this notation the action \p{PSTMact} and its gauge symmetries \p{PTSMgaugetr} read
\bea
S&=& \int d^6 x \left[ \partial^{(\mu\sigma}B_\sigma{}^{\nu)} \partial_{(\mu\rho}B_{\nu)}{}^\rho -2 \partial^{(\mu\sigma}B_\sigma{}^{\nu)} \partial_{(\mu\rho}z R_{\nu)}{}^\rho + \partial_{(\mu\rho}z R_{\nu)}{}^\rho \, \partial^{(\mu\sigma}z R_\sigma{}^{\nu)} \right], \label{PSTMact2}\\
&&1)\;\; \delta_f B_\alpha{}^\beta = \partial_{\alpha\gamma}f^{\beta\gamma} - \frac{1}{4}\delta_\alpha^\beta \partial_{\mu\nu}f^{\mu\nu}, \;\; \delta_f R_\alpha{}^\beta=0, \;\; \delta_f z=0, \label{trans1} \\
&&2)\;\; \delta_a B_\alpha{}^\beta = \partial_{\alpha\gamma }z\,a^{\beta\gamma} - \frac{1}{4}\delta_\alpha^\beta \partial_{\mu\nu }z\, a^{\mu\nu}, \;\; \delta_a R_\alpha{}^\beta = -\partial_{\alpha\gamma}a^{\beta\gamma} + \frac{1}{4}\delta_\alpha^\beta \partial_{\mu\nu}a^{\mu\nu}, \;\; \delta_a z=0,\label{trans2} \\
&&3)\;\; \delta_b R_\alpha{}^\beta = \partial_{\alpha\gamma }z\,b^{\beta\gamma} - \frac{1}{4}\delta_\alpha^\beta \partial_{\mu\nu }z\, b^{\mu\nu}, \;\; \delta_b B_\alpha{}^\beta=0, \;\; \delta_b z=0, \label{trans3} \\
&&4)\;\; \delta_\lambda z = \lambda, \;\; \delta_\lambda B_\alpha{}^\beta = \lambda R_\alpha{}^\beta, \;\; \delta_\lambda R_\alpha{}^\beta = -4\lambda \frac{\partial^{\beta\rho}z}{\partial_{\mu\nu}z\, \partial^{\mu\nu}z}\partial_{(\alpha\sigma}R_{\rho)}{}^\sigma. \label{trans4}
\eea
Here spinor indices take values $\alpha =1\ldots 4$. In this notation vector $a_M$ corresponds to an antisymmetric object $a_{[\mu\nu]}$, a 2-form $B_{MN}$ is a traceless matrix $B_\alpha{}^\beta$, and objects with two symmetric lower or upper indices are self-dual or anti-self-dual 3-forms. Antisymmetric pairs of spinor indices can be raised and lowered by completely antisymmetric symbols $\epsilon_{\alpha\beta\mu\nu}$, $\epsilon^{\alpha\beta\mu\nu}$, $\epsilon_{1234}=\epsilon^{1234}=1$, $\epsilon_{\alpha\beta\mu\nu}\epsilon^{\alpha\beta\mu\nu}=24$:
\be\label{raisingspinor}
a^{[\alpha\beta]} = \frac{1}{2}\epsilon^{\alpha\beta\mu\nu}a_{[\mu\nu]}, \;\; a_{[\alpha\beta]} = \frac{1}{2}\epsilon_{\alpha\beta\mu\nu}a^{[\mu\nu]},
\ee
so that one can easily form scalar product of $d=6$ vectors. From \p{raisingspinor} it follows that for one and two vectors
\be\label{vectprop}
a_{\alpha\gamma}a^{\beta\gamma} = \frac{1}{4}\delta_\alpha^\beta\, a_{\mu\nu}a^{\mu\nu}, \;\; a_{\alpha\gamma}b^{\beta\gamma} = - a^{\beta\gamma}b_{\alpha\gamma} + \frac{1}{2}\delta_\alpha^\beta\,a_{\mu\nu}b^{\mu\nu}.
\ee
These properties will be widely used further in this paper.

The goal of this work is to find the superfield action that in the bosonic limit reproduces \p{PSTMact2} with gauge symmetries that extend \p{trans1}, \p{trans2}, \p{trans3}, \p{trans4}.

\setcounter{equation}0
\section{Harmonic superspace and tensor multiplets}
To reproduce the action \p{PSTMact2} using superspace techniques we introduce the $N=(1,0)$, $d=6$ harmonic superspace \cite{d6HS,buchbinder} and consider all the fields involved as the components of harmonic superfields.
The $N=(1,0)$, $d=6$ harmonic superspace in the standard basis can be parameterized by the usual spacetime coordinates $x^{[\alpha\beta]}$, odd coordinates $\theta^{\alpha}_i$ and the harmonics $u^{+i}$, $u^{-j}$, $i,j=1,2$, which satisfy
\be\label{harmdef}
u^{+i}u^{-j}\epsilon_{ij} =1, \;\; \epsilon_{ij} = -\epsilon_{ji}, \;\; \epsilon_{12}=1
\ee
and parameterize unit 2-sphere. The covariant derivatives with respect to these coordinates are given by the relations
\be\label{covdersusual}
\partial_{\alpha\beta}, \;\; D^i_\alpha =  \frac{\partial}{\partial \theta^{\alpha}_i} -\im \theta^{i\beta}\partial_{\alpha\beta}, \;\; \partial^{++} = u^{+i}\frac{\partial}{\partial u^{-i}}, \;\; \partial^{--} = u^{-i}\frac{\partial}{\partial u^{+i}}, \;\; \partial_0 =  u^{+i}\frac{\partial}{\partial u^{+i}} -u^{-i}\frac{\partial}{\partial u^{-i}}.
\ee
Superfields defined on this superspace have to possess definite charge: $\partial_0 f^{q} = q f^{q}$. This reflects the fact that harmonics describe $S^2 = SU(2)/U(1)$, not whole $SU(2)$. Thus harmonic superfields are power series in harmonics with properly balanced charges. For example, for positive charge $q$
\be\label{harmsfexp}
f^{q}(x,\theta,u) = \sum^{\infty}_{n=0}f_{(i_1 \, \ldots \, i_{q+n}, j_1, \ldots, j_n)}(x,\theta) u^{+ i_1}\ldots u^{+i_{q+n}}  u^{-j_1} \ldots u^{-j_n}.
\ee
For $f^{q}$ with negative charge, roles of $u^{+i}$ and $u^{-i}$ are inverted. If net charge is zero, a harmonic-independent part may be present.

Important reason of using the harmonic superspace formalism is the ability to perform change of the coordinates and pass to the so-called analytic basis with the new coordinates being $x_{(a)}^{\alpha\beta}$, $\theta^{+\alpha}$, $\theta^{-\alpha}$, $u^{+i}$, $u^{-j}$. Let us omit explicit relations between coordinates in these bases and write down only the covariant derivatives:
\bea\label{covdersanalyt}
D^{+}_\alpha = \frac{\partial}{\partial \theta^{-\alpha}}, \;\; D^{-}_\alpha = -\frac{\partial}{\partial \theta^{+\alpha}} -2\im \theta^{-\beta}\partial_{\alpha\beta}, \;\;  D_0 = \partial_0 + \theta^{+\gamma}\frac{\partial}{\partial \theta^{+\gamma}} -\theta^{-\gamma}\frac{\partial}{\partial \theta^{-\gamma}},  \nn \\
D^{++} = \partial^{++} + \im \theta^{+\alpha}\theta^{+\beta}\partial_{\alpha\beta} + \theta^{+\gamma}\frac{\partial}{\partial \theta^{-\gamma}}, \;\; D^{--}=\partial^{--} + \im \theta^{-\alpha}\theta^{-\beta}\partial_{\alpha\beta} + \theta^{-\gamma}\frac{\partial}{\partial \theta^{+\gamma}}.
\eea
They satisfy the (anti)commutation relations
\bea\label{hsderscom}
\big\{ D^{+}_\alpha, D^{-}_\beta  \big\}=2\im \partial_{\alpha\beta}, \;\; \big\{ D^{+}_\alpha, D^{+}_{\beta}\big\}=0, \;\; \big\{ D^{-}_\alpha, D^{-}_{\beta}\big\}=0, \nn \\
\big[ D^{++},D^{--}  \big]=D_0, \;\; \big[ D_0, D^{++} \big]=2D^{++}, \;\; \big[ D_0, D^{--} \big]=-2D^{--}, \nn \\
\big[ D^{++},D^{+}_\alpha   \big] =0, \;\; \big[ D^{--},D^{+}_\alpha   \big] =D^{-}_\alpha, \;\; \big[ D_0,D^{+}_\alpha   \big] = D^{+}_\alpha, \nn \\
\big[ D^{++},D^{-}_\alpha \big] =D^{+}_\alpha,  \;\; \big[ D^{--},D^{-}_\alpha   \big] =0, \;\; \big[ D_0,D^{-}_\alpha   \big] = -D^{-}_\alpha.
\eea
In the analytic basis $x_{(a)}^{\alpha\beta}$, $\theta^{+\alpha}$, $u^{+i}$, $u^{-j}$ form a subspace, invariant with respect to $N=(1,0)$, $d=6$ supersymmetry transformations, and the covariant derivative $D^{+}_\alpha$ involves differentiation with respect to $\theta^{-\alpha}$ only. This property allows to consider so-called analytic superfields, which do not depend on $\theta^{-\alpha}$ and can be integrated over the analytic subspace $x_{(a)}^{\alpha\beta}$, $\theta^{+\alpha}$, $u^{+i}$, $u^{-j}$. These superfields play crucial role in the description of the $N=(1,0)$, $d=6$ Yang-Mills theory and matter, as the Yang-Mills prepotential $V^{++}$ and the hypermultiplet superfield $q^{+}_a$ are unconstrained analytic superfields \cite{d6HS}. The situation with tensor multiplet is more complicated, which we discuss below. However, the action we are going to construct will be an integral over analytic subspace.

Integration over Grassmann coordinates in the analytic superspace is defined as $\int d^4\theta^{(-)}\theta^{+4} = 1$ \footnote{We define $\big(\theta^{+3}\big)_{\alpha} = \frac{1}{6}\epsilon_{\alpha\mu\nu\lambda}\theta^{+\mu} \theta^{+\nu} \theta^{+\lambda}$, $ \theta^{+4} = -\frac{1}{24}\epsilon_{\alpha\beta\mu\nu} \theta^{+\alpha}\theta^{+\beta}\theta^{+\mu}\theta^{+\nu}$, $\theta^{+\alpha}\big( \theta^{+3} \big)_\beta = -\delta_\beta^\alpha \, \theta^{+4}$}. Integration over harmonics can be performed using the rules \cite{HS}
\be\label{harmintegr}
\int du \, 1 =1, \;\; \int du u^{+}_{(i_1} \ldots u^{+}_{i_n} u^{-}_{j_1} \ldots u^{-}_{j_m)} =0, \;\; m\; \mbox{or}\; n\neq 0.
\ee

$N=(1,0)$ tensor multiplet, both in conventional and harmonic superspaces, can be described in two different ways \cite{sokatchev}.

The first way is to introduce the real superfield $\Phi$ that satisfies the constraint $D_\alpha^{(i}D^{j)}_\beta \Phi =0$. This constraint reduces the component content of $\Phi$ to $\phi = \Phi|_{\theta\rightarrow 0}$, $\chi^i_\alpha = D^i_\alpha \Phi|_{\theta\rightarrow 0}$, $G_{(\alpha\beta)} = D^i_{(\alpha} D_{i\beta)}\Phi|_{\theta\rightarrow 0}$ and puts these components on shell:
\be\label{tmfreeeoms1}
\partial^{\alpha\beta}\partial_{\alpha\beta} \phi =0, \;\; \partial^{\alpha\beta}\chi^i_\beta =0, \;\; \partial^{\alpha\beta}G_{(\beta\gamma)} =0.
\ee
The field $G_{(\alpha\beta)}$ is usually assumed to be the self-dual part of the field strength of some 2-form: $G_{(\alpha\beta)} = \partial_{(\alpha\gamma}B_{\beta)}^\gamma$.

As the components involve just one $SU(2)$ spinor, the constraint $D_\alpha^{(i}D^{j)}_\beta \Phi =0$ is equivalent to harmonic ones $D^{+}_\alpha D^{+}_\beta \Phi =0$ and $D^{++}\Phi=0$. One, therefore, can consider the harmonic superfield $\Phi(x, \theta^+, \theta^-, u)$ constrained by $D^{+}_\alpha D^{+}_\beta \Phi =0$ and treat $D^{++}\Phi=0$ as the equation of motion. The constraint $D^{+}_\alpha D^{+}_\beta \Phi =0$ implies that $\Phi$ is linear in $\theta^-$: $\Phi = b(x, \theta^+,u) + \theta^{-\alpha}b^+_\alpha(x, \theta^+,u)$.

Alternatively, one can introduce the superfield $X^{i\alpha}$ and subject it to the condition $D^{(i}_\alpha X^{j)\beta} = \frac{1}{4}\delta_\alpha^\beta D^{(i}_\gamma X^{j)\gamma}$, which looks like the condition on the vector multiplet superfield strength but without additional restriction on the scalar component. Unlike the superfield $\Phi$, $X^{i\alpha}$ contains the 2-form potential explicitly and has nontrivial gauge transformation law $\delta X^{i\alpha} = W^{i\alpha}$, where $W^{i\alpha}$ is an infinitesimal abelian vector multiplet field strength. As $D^i_\alpha W^\alpha_i=0$, this gauge transformation preserves the constraint on $X^{i\alpha}$. Just like the previous formulation, this one is on-shell and selects bosons $q = D^i_\alpha X^\alpha_i$, $B_\alpha{}^\beta = D^i_\alpha X_i^\beta$ and the fermion $\psi^i_\alpha = D^i_\alpha D^j_\beta X^{\beta}_j$ as only independent dynamical components, while some first components are purely gauge degrees of freedom. Fields $q$, $B_\alpha{}^\beta$ and $\psi^i_\alpha$ are subjected to the equations of motion
\be\label{tmfreeeoms2}
\partial^{\alpha\beta}\partial_{\alpha\beta} q =0, \;\; \partial^{\alpha\beta}\psi^i_\beta =0, \;\; \partial^{(\alpha\gamma}B_\gamma{}^{\beta)} =0,
\ee
which include self-duality equation. In the harmonic superspace, the constraint on $X^{i\alpha}$ is equivalent to $D^+_\alpha X^{+\beta} = \frac{1}{4}\delta_\alpha^\beta D^+_\gamma X^{+\gamma}$ and $D^{++}X^{+\alpha}=0$. Again, one can treat $D^{++}X^{+\alpha}=0$ as an equation of motion, making the other constraint off-shell. The condition $D^+_\alpha X^{+\beta} \sim \delta_\alpha^\beta$ implies that superfield $X^{+\alpha}$ has structure $X^{+\alpha} = v^{+\alpha}(x,\theta^{+},u) + \theta^{-\alpha} v^{++}(x,\theta^{+},u)$.

As both formulations involve superfields that depend on $\theta^{-\alpha}$, construction of integrals over analytic superspace that could serve as the action functional is non-trivial. However, using the constraints $D^+_\alpha D^+_\beta \Phi = 0$ and $D^+_\alpha X^{+\beta} = \frac{1}{4}\delta_\alpha^\beta D^+_\gamma X^{+\gamma}$, it is possible to show that
\bea\label{lagrdouble}
D^+_\alpha \left[ D^{+}_\beta \Phi\, X^{+\beta} + \frac{1}{4}\Phi\,  D^{+}_\beta X^{+\beta}   \right]= 0\;\; \mbox{and, as}\;\; \big[ D^{++},D^+_\alpha \big] =0, \nn \\
D^+_\alpha \left[ D^{+}_\beta \Phi\, D^{++}X^{+\beta} + \frac{1}{4}\Phi\, D^{++}D^{+}_\beta X^{+\beta}   \right]= 0.
\eea
The second invariant has the right charge and dimension and can serve as the superfield Lagrangian \cite{sokatchev,buchbinder}:
\be\label{buchact}
S = \int d^6 x d^4\theta^{-}du \Big[ D^{+}_\alpha \Phi D^{++}X^{+\alpha} + \frac{1}{4}\Phi D^{++}D^{+}_\alpha X^{+\alpha}  \Big].
\ee
As the superfields $\Phi$, $X^{+\alpha}$ are not analytic, to find the equations of motion it is necessary to introduce unconstrained prepotentials for each supermultiplet \footnote{We define $\big(D^{+3}\big)^{\alpha} = - \frac{1}{6}\epsilon^{\alpha\mu\nu\lambda}D^+_\mu D^+_\nu D^+_\lambda$, $D^{+4} = -\frac{1}{24}\epsilon^{\alpha\beta\mu\nu}D^+_\alpha D^+_\beta D^+_\mu D^+_\nu $, $D_\beta D^{+3\alpha} = \delta_\beta^\alpha D^{+4}$} $\Phi = \big(D^{+3}\big)^{\alpha}\Phi^{-3}_\alpha$, $X^{+\alpha} = \big(D^{+3}\big)^{\alpha}X^{--}$ , rewrite the action as the integral over the whole superspace and vary with respect to prepotentials. The equations of motion are, actually, the right ones $D^{++}\Phi =0$, $D^{++}X^{+\alpha} =0$. The obvious disadvantage of this action is the fact that it describes two physical multiplets, not one. The idea of how to relate these two multiplets was given by Sokatchev \cite{sokatchev} and involves using the supergravity multiplet fields as Lagrangian multipliers, though this approach leads to a condition $q\phi =1$.

\setcounter{equation}0
\section{Extending the PST action}
Not satisfied with the results of \cite{sokatchev}, we decided to follow different approach and construct the superfield action that would involve only one physical multiplet and extend the Pasti-Sorokin-Tonin action \cite{PST} in formulation suggested by \cite{mkrtchyan}. Thus, as the first step, we should reproduce the standard bilinear kinetic term of the 2-form gauge field. As the $\Phi$ superfield does not involve the 2-form explicitly, one should construct the bilinear functional of $X^{+\alpha}$ which would be an integral over analytic superspace. The idea how to find one is to observe that the analog of the superfield $\Phi$ can be constructed in terms of superfield $X^{+\alpha}$. Indeed, taking into account dimensions and charges, one can suggest that
\be\label{PhiX}
\Phi\big[ X \big] = a_1 D^{--}D^+_\alpha X^{+\alpha}  + a_2 D^-_\alpha X^{+\alpha}.
\ee
This expression satisfies $D^+_\alpha D^+_\beta \Phi\big[ X \big] =0$ if $a_2 =-2a_1$. Thus we could take $a_1 =1$, $a_2 =-2$ and consider the kinetic term functional as
\be\label{S1}
S_1 = \int d^6 x d^4\theta^{(-)}du \Big[ D^{+}_\alpha  \Phi\big[ X \big] D^{++}X^{+\alpha} + \frac{1}{4}\Phi\big[ X \big] D^{++}D^{+}_\alpha X^{+\alpha}  \Big].
\ee
Quick check can be performed by substituting $X^{+\alpha} \approx \theta^{+\beta} \big( \delta_\beta^\alpha q + B_\beta{}^\alpha \big) + \theta^{-\alpha}\theta^{+\mu}\theta^{+\nu}a_{\mu\nu}$, where fermionic and charged bosonic components, as well as harmonic dependence, were neglected. Extracting $\theta^{+4}$ component of the superfield Lagrangian \p{S1} and taking into account that harmonic integration becomes trivial, one obtains
\be\label{S1approx1}
S_1 \approx \int d^6 x \big[  8\big( a_{\alpha\beta} - \im \partial_{[\alpha\rho}B_{\beta]}{}^{\rho}- \im \partial_{\alpha\beta} q \big)\big( a^{\alpha\beta} + \im \partial^{[\alpha\sigma}B_{\sigma}{}^{\beta]}+ \im \partial_{\alpha\beta} q \big) -8 \partial_{(\alpha\rho} B_{\beta)}{}^{\rho}\partial^{(\alpha\sigma} B_{\sigma}{}^{\beta)} -16\im a^{\mu\nu}\partial_{\mu\nu} q\big].
\ee
Removing auxiliary field $a_{\alpha\beta}$ by its equation of motion, one obtains
\be\label{S1approx2}
S_1 \approx \int d^6 x \big[ -8 \partial_{(\alpha\rho} B_{\beta)}{}^{\rho}\partial^{(\alpha\sigma} B_{\sigma}{}^{\beta)} + 16 \partial^{\alpha\beta} q \partial_{\alpha\beta} q  \big].
\ee
Thus $S_1$ contains the correct kinetic terms for both scalar and tensor fields. However, $S_1$ can not be a correct action for the tensor multiplet on its own. Equation of motion, obtained by varying \p{S1} with respect to prepotential $X^{--}$, reads $D^{++}\Phi[X] =0$. It does not contain the self-duality equation for the field strength of $B_\alpha{}^\beta$. Moreover, it does not show that $B_\alpha{}^\beta$ does not depend on harmonics. Instead, it implies equation $\partial_{(\alpha\gamma}\partial^{++}B_{\beta)}{}{^\gamma} =0$, and, therefore, $B_\alpha{}^\beta$ contains infinite tower of anti-self-dual fields in its harmonic expansion. All these results are consistent with the already known one that it is not possible to construct an action of self-dual field that does not contain supplementary fields. Therefore, more terms should be added to the action to remove these unwanted terms. As will be shown later, parts of the Lagrangian needed to obtain the PST action perfectly play this role.

The second term that should be added to the action should be linear in the physical fields and bilinear in auxiliary ones. Moreover, it should contain a coupling to the anti-self-dual part of $B$ field strength, $\partial^{(\alpha\gamma}B_\gamma{}^{\beta)}$. As $\Phi\big[ X \big]$ does not contain it, the $X^{+\alpha}$ superfield could enter this cubic term only as $D^{++}X^{+\alpha}$ or $D^{++}D^+_\alpha X^{+\alpha}$. It can be shown that analytic coupling, linear in $D^{++}X$, has to take the form
\be\label{S2candfail}
D^{+}_\beta H\,D^{++} X^{+\beta} + \frac{1}{4}H\,D^{++}  D^{+}_\beta X^{+\beta}, \;\; D^+_\alpha D^{+}_\beta H =0.
\ee
Superfunction $H$ should be bilinear in auxiliary superfields. Though the field content of the polynomial action requires one scalar and one tensor field which nicely matches the components of a tensor multiplet $Y^{+\alpha}$, one can show by analyzing the appropriate ansatz that it is not possible to construct $H$ that is bilinear in $Y^{+\alpha}$ and satisfies $D^+_\alpha D^{+}_\beta H =0$. To circumvent this, we introduce an analytic superfield $Z$ which first component is the auxiliary scalar. The proper coupling, therefore, should be
\be\label{S2}
S_2 = \int d^6 x d^4\theta^{(-)}du \Big[ D^{+}_\alpha  H\big[ Z,Y \big] D^{++}X^{+\alpha} + \frac{1}{4}H\big[Z,Y \big] D^{++}D^{+}_\alpha X^{+\alpha}  \Big], \;\; D^+_\alpha D^+_\beta H[Z,Y] =0.
\ee
As $H\big[ Z,Y \big]$ is linear in $Z$ and $Y$, it could only be a combination of $\Phi\big[ Z\cdot Y^{+\alpha}  \big]$ and $Z \Phi\big[ Y^{+\alpha}  \big]$. To avoid excessive derivatives on one of the fields, one should take
\be\label{HZY}
H\big[ Z,Y \big] =\Phi\big[ Z\cdot Y^{+\alpha}  \big]- Z \Phi\big[ Y^{+\alpha}  \big] = D^{--}Z \,D^+_\alpha Y^{+\alpha} - 2 D^{-}_\alpha Z\, Y^{+\alpha}.
\ee
The third term should be bilinear in $Z$ and $Y$. It can also be modelled analogously to \p{S2}, replacing $D^{++}X^{+\alpha}$ with $D^{++}Z Y^{+\alpha}$, which does not spoil analyticity and does not introduce excessive derivatives:
\be\label{S3}
S_3 = \int d^6 x d^4\theta^{(-)}du D^{++}Z\Big[ D^{+}_\alpha  H\big[ Z,Y \big] Y^{+\alpha} + \frac{1}{4}H\big[Z,Y \big] D^{+}_\alpha Y^{+\alpha}  \Big].
\ee
Combination $S_1 + k_2 S_2 + k_3 S_3$ contains the complete PST action in form suggested by Mkrtchyan \cite{mkrtchyan}. This can be quickly checked by inserting superfields with neglected fermionic and charged bosonic components:
\be
X^{+\alpha} \approx \theta^{+\beta} \big( \delta_\beta^\alpha q + B_\beta{}^\alpha \big) + \theta^{-\alpha}\theta^{+\mu}\theta^{+\nu}a_{\mu\nu}, \;\; Y^{+\alpha} \approx \theta^{+\beta} \big( \delta_\beta^\alpha c + R_\beta{}^\alpha \big) + \theta^{-\alpha}\theta^{+\mu}\theta^{+\nu}b_{\mu\nu}, \;\; Z \approx z \;\; \label{XYZsimple} \\
\ee
Performing $\theta$-integration and excluding auxiliary fields by their equations of motion, one finds
\bea
{\cal L} \approx -2\partial_{\alpha\beta}q\partial^{\alpha\beta}q +\partial_{(\alpha\rho}B_{\beta)}{}^{\rho} \partial^{(\alpha\sigma}B_\sigma{}^{\beta)}+k_1 \partial_{(\alpha \rho}z\, R_{\beta)}{}^\rho \, \partial^{(\alpha\sigma}B_\sigma{}^{\beta)}+\nn \\ + \frac{1}{4} k_1^2 \partial_{[\alpha \rho}z\, R_{\beta]}{}^\rho \partial^{[\alpha \sigma}z\, R_\sigma{}^{\beta]}-k_2 \partial_{\alpha\beta}z\partial^{\mu\nu}z \,R_\mu{}^\alpha R_\nu{}^\beta + \big( -\frac{1}{4}k_1^2 -k_2  \big)c^2 \partial_{\alpha\beta}z\partial^{\alpha\beta}z. \label{Lapprox}
\eea
Therefore, the desired bosonic action \p{PSTMact2} can be recovered if $k_2 =-2$, $k_3 =1$. This is not the end of the story, however, as we also need to find the gauge symmetries of our action and show that it does not contain excessive degrees of freedom, which could appear in the harmonic expansions. As the solution to the second problem strongly depends on the results of solving the first one, let us reconstruct the gauge symmetries that generalize \p{trans1}, \p{trans2}, \p{trans3}, \p{trans4} and show that action $-8 S = S_1 -2S_2 + S_3$ is invariant with respect to them after minor modifications.

\setcounter{equation}0
\section{Gauge symmetries}
The Pasti-Sorokin-Tonin action in its original and polynomial forms possesses a set of gauge symmetries which are needed to show that this action produces self-dual equation of motion for 2-form field and does not introduce new degrees of freedom. It should be expected that these symmetries extend to the supersymmetric action
\bea\label{semifinalact}
-8 S[X,Y,Z]&=&  \int d^6 x d^4\theta^{-} du \Big[ D^{+}_\beta \Phi\big[ X\big]  \,D^{++}X^{+\beta} + \frac{1}{4}\Phi\big[ X\big] \,D^{++} D^{+}_\beta X^{+\beta} -  \\
&&-2 \Big(  D^{+}_\beta H\big[ Z,Y\big]  \,D^{++}X^{+\beta} + \frac{1}{4}H\big[ Z,Y\big] \,D^{++} D^{+}_\beta X^{+\beta}   \Big) + \nn \\
&&+ D^{++}Z \Big( D^{+}_\beta H\big[ Z,Y\big]  \,Y^{+\beta} + \frac{1}{4}H\big[ Z,Y\big] \, D^{+}_\beta Y^{+\beta}\Big) \Big].\nn
\eea
The first symmetry, a shift of the 2-form physical field by a differential of one-form field, is built into the structure of tensor multiplet and is realized as $\delta X^{+\alpha} =W^{+\alpha}$, where $W^{+\alpha}$ is an infinitesimal abelian vector multiplet strength, satisfying \cite{buchbinder}
\be\label{vectmultprop}
D^+_\alpha W^{+\beta} =\frac{1}{4}\delta_\alpha^\beta \, D^+_\gamma W^{+\gamma}, \;\; D^{++}W^{+\alpha} =0, \;\; D^{--}D^+_\alpha W^{+\alpha} -2 D^-_\alpha W^{+\alpha} =0.
\ee
Looking at the structure of the second transformation, which involves shift of $R$ 2-form by differential of 1-form, one can guess that $Y^{+\alpha}$ should have transformation law $\delta Y^{+\alpha} =W^{+\alpha}$. The $X^{+\alpha}$ transformation then should be taken as $\delta X^{+\alpha} =Z W^{+\alpha}$ to make whole action invariant. It could be checked that actually \p{semifinalact} is invariant with respect to infinite set of transformations
\bea\label{deltaZW}
&&\delta X^{+\alpha} = \frac{Z^{n+1}}{n+1}W^{+\alpha}, \;\; \delta Y^{+\alpha} = Z^n W^{+\alpha} \;\; \Rightarrow \nn \\
-8 \delta S\big[X,Y,Z\big]&=&  \int d^6 x d^4\theta^{-} du \Big[2 D^{+}_\beta H\big[ Z,W\big]  \,D^{++}X^{+\beta} + \frac{1}{2}H\big[ Z,W\big] \,D^{++} D^{+}_\beta X^{+\beta} + \nn \\
&&-2 \Big(  D^{+}_\beta H\big[ Z,Y\big]  \,D^{++}Z W^{+\beta} + \frac{1}{4}H\big[ Z,Y\big] \,D^{++} Z D^{+}_\beta W^{+\beta}   \Big)-  \\
&&-2\Big(  D^{+}_\beta H\big[ Z,W\big]  \,D^{++}X^{+\beta} + \frac{1}{4}H\big[ Z,W\big] \,D^{++} D^{+}_\beta X^{+\beta}   \Big)
 + \nn \\
&&+2 D^{++}Z \big( D^{+}_\beta H\big[ Z,W\big]  \,Y^{+\beta} + \frac{1}{4}H\big[ Z,W\big] \, D^{+}_\beta Y^{+\beta}\Big) \Big]Z^n=0.\nn
\eea
When checking the invariance one should use identities
\be\label{deltaZW2}
\delta\Phi\big[ X \big] =Z^n H\big[ Z,W  \big], \;\; \delta D^{++}X^{+\alpha} = Z^n\, D^{++}Z\, W^{+\alpha},
\ee
which follow from the properties of the abelian vector multiplet \p{vectmultprop}. Combining transformations \p{deltaZW} with different parameters $W^{+\alpha}$, one can reproduce all three transformations of the basic fields that leave $Z$ invariant. Taking $W_1$, $W_2$, $W_3$ with following tensor components
\bea
W_1^{+\alpha} = \ldots + \theta^{+\rho} \Big( \partial_{\rho\gamma}b^{\alpha\gamma} - \frac{1}{4} \delta_\rho^\alpha \partial_{\mu\nu}b^{\mu\nu} \Big) + \ldots, \nn \\
W_2^{+\alpha} = \ldots + \theta^{+\rho} \Big( \partial_{\rho\gamma}\big(z b^{\alpha\gamma}\big) - \frac{1}{4} \delta_\rho^\alpha \partial_{\mu\nu}\big(z b^{\mu\nu}\big) \Big) + \ldots,  \\
W_3^{+\alpha} = \ldots +  \theta^{+\rho} \Big( \partial_{\rho\gamma}\big(z^2 b^{\alpha\gamma}\big) - \frac{1}{4} \delta_\rho^\alpha \partial_{\mu\nu}\big(z^2 b^{\mu\nu}\big) \Big) + \ldots, \nn
\eea
one can reproduce the third transformation \p{trans3} via
\be\label{btrans}
\delta_b Y^{+\alpha} = W_2^{+\alpha}-Z W_1^{+\alpha}, \;\; \delta_b X^{+\alpha} = Z W_2^{+\alpha} -\frac{Z^2}{2}W_1^{+\alpha}-\frac{1}{2}W_3^{+\alpha}.
\ee
In the latter transformation terms with $b^{\mu\nu}$ cancel entirely, as should be expected. It is not yet known how many transformations \p{deltaZW} are truly independent.

The fourth symmetry \p{trans4}, which involves shift of $z$ by arbitrary function, has to be generalized in different way. Moreover, it is required to introduce semi-trivial modification of the action to reconstruct it. Let us take
\be\label{modact}
-8 S_{mod}= -8 S + \int d^6 x d^4 \theta^{-}du \, M \big( D^{++}D^+_\gamma X^{+\gamma} - D^{++}Z  D^+_\gamma Y^{+\gamma} + N^{+4} \big),
\ee
where $M$ and $N^{+4}$ are some analytic superfields. Then one can write down variation of \p{modact}, assuming that $Z$ acquires a shift by an unconstrained infinitesimal analytic superfield $\Lambda$:
\bea\label{delta4}
&&-8 \delta_\Lambda S_{mod}=  \int d^6 x d^4\theta^{-} du \Big[2 D^{+}_\beta \Phi\big[ \delta_\Lambda X\big]  \,D^{++}X^{+\beta} + \frac{1}{2}\Phi\big[ \delta_\Lambda X\big] \,D^{++} D^{+}_\beta X^{+\beta} + \nn \\
&&-2\Big(  D^{+}_\beta \delta_\Lambda H\big[ Z,Y \big]  \,D^{++}X^{+\beta} + \frac{1}{4}\delta_\Lambda H\big[ Z, Y \big] \,D^{++} D^{+}_\beta X^{+\beta}   \Big) -2 \Big(  D^{+}_\beta H\big[ Z,Y\big]  \,D^{++}\delta_\Lambda X^{+\beta}  + \nn \\
&&+ \frac{1}{4}H\big[ Z,Y\big] \,D^{++} D^{+}_\beta \delta_\Lambda X^{+\beta}   \Big)+ D^{++}\Lambda \big( D^{+}_\beta H\big[ Z,Y\big]  \,Y^{+\beta} + \frac{1}{4}H\big[ Z,Y\big] \, D^{+}_\beta Y^{+\beta}\Big)  +\nn \\
&&+ D^{++}Z\big(  D^{+}_\beta H\big[ Z,Y\big]  \,\delta_\Lambda Y^{+\beta} + \frac{1}{4} H\big[ Z ,Y\big] \, D^{+}_\beta \delta_\Lambda Y^{+\beta}\Big) +\nn \\
&&+ D^{++}Z \big( D^{+}_\beta \delta_\Lambda H\big[ Z,Y\big]  \,Y^{+\beta} + \frac{1}{4}\delta_\Lambda H\big[ Z,Y\big] \, D^{+}_\beta Y^{+\beta}\Big) -  \\ &&-
D^+_\alpha X^{+\alpha} D^{++}\delta_\Lambda M + M\, D^{++}D^+_\alpha \delta_\Lambda  X^{+\alpha} -\delta_\Lambda M\, D^{++}Z D^+_\gamma Y^{+\gamma} -\nn \\
&& - D^{++}\Lambda \, M\, D^+_\gamma Y^{+\gamma} - D^{++}Z \, M\, D^+_\gamma\delta_\Lambda Y^{+\gamma} + \delta_\Lambda M \, N^{+4} + M\, \delta_\Lambda N^{+4} \Big].\nn
\eea
As transformations of all the fields in \p{trans4} do not depend on $B_\alpha{}^\beta$, one can assume that the superfield ones do not involve $X^{+\alpha}$. Therefore, one can represent all the terms with $X^{+\alpha}$ as an integral
\be\label{delta41}
-2\int d^6 x d^8 \theta du \big[ X^{--}D^{++}\big( \Phi\big[ \delta_\Lambda X\big] - \delta_\Lambda H\big[Z,Y  \big] +2 \delta_\Lambda M \big)   \big]
\ee
and deduce that the transformations should satisfy relation $D^{++}\big( \Phi\big[ \delta_\Lambda X\big] - \delta_\Lambda H\big[Z,Y  \big] +2 \delta_\Lambda\, M\big) =0$. As it looks like variation of equation of motion, one can strengthen this condition to
\be\label{delta42}
\Phi\big[ \delta_\Lambda X\big] - \delta_\Lambda H\big[Z,Y  \big] +2 \delta_\Lambda M=0.
\ee
As $\delta_\Lambda H\big[Z,Y  \big] = H\big[ \Lambda, Y \big] + H\big[ Z, \delta_\Lambda Y \big]$, one can use nonanalytic part of \p{delta42} to find $\delta_\Lambda Y^{+\alpha}$ and then find $\delta_\Lambda M$ from analytic part:
\bea\label{delta421}
D^+_\alpha \Phi\big[ \delta_\Lambda X\big] -D^+_\alpha H\big[Z, \delta_\Lambda Y  \big] - D^+_\alpha H\big[\Lambda, Y  \big] =0 \;\; \Rightarrow \nn \\
4\im \partial_{\alpha\beta}Z \, \delta_\Lambda Y^{+\beta} +\frac{1}{2} D^-_\alpha Z\, \delta_\Lambda D^+_\beta Y^{+\beta} - D^+_\alpha H\big[\Lambda, Y  \big] +D^+_\alpha \Phi\big[ \delta_\Lambda X\big]=0 \;\; \Rightarrow  \nn \\
 \delta_\Lambda Y^{+\alpha} =-\im  \frac{\partial^{\alpha\beta}Z}{\partial_{\mu\nu}Z \, \partial^{\mu\nu}Z}\left[ \frac{1}{2}  D^-_\beta Z\, \delta_\Lambda D^+_\gamma Y^{+\gamma} + D^+_\beta \Phi\big[ \delta_\Lambda X\big] -D^+_\beta H\big[ \Lambda,Y \big] \right],\nn \\
\delta_\Lambda M = -\frac{1}{2} \left[ \big( \Phi\big[ \delta_\Lambda X\big]  -  H\big[\Lambda, Y  \big] \big) - 2\im \frac{D^-_\beta Z\, \partial^{\beta\gamma}Z \, D^+_\gamma \big( \Phi\big[ \delta_\Lambda X\big]  -  H\big[\Lambda, Y  \big] \big)}{\partial_{\mu\nu}Z \, \partial^{\mu\nu}Z}  - \widehat{D^{--}Z }\delta_\Lambda D^+_\gamma Y^{+\gamma}\right], \nn \\
\mbox{where}\;\; \widehat{D^{--}Z } = D^{--}Z + \im \frac{\partial^{\alpha\beta}Z \, D^-_\alpha Z\, D^-_\beta Z}{\partial_{\mu\nu}Z \, \partial^{\mu\nu}Z}, \;\; D^+_\alpha \widehat{D^{--}Z }=0.
\eea
Note that $\delta_\Lambda D^+_\alpha Y^{+\alpha} $ remains independent, and equation \p{delta42} effectively determines analytic part of $\delta_\Lambda Y^{+\alpha}$.

Variations \p{delta421} can be substituted back to \p{delta4} to simplify it. However, as
\bea\label{delta43}
D^{+}_\beta H\big[ Z,Y\big]  \,\delta_\Lambda Y^{+\beta} + \frac{1}{4}\delta_\Lambda  H\big[ Z ,Y\big] \, D^{+}_\beta \delta_\Lambda Y^{+\beta} = D^{+}_\beta H\big[ Z,\delta_\Lambda Y\big]  \, Y^{+\beta} + \frac{1}{4} H\big[ Z , \delta_\Lambda Y\big] \, D^{+}_\beta Y^{+\beta} \Rightarrow \nn \\
\delta_\Lambda \big( D^{+}_\beta H\big[ Z,Y\big]  \, Y^{+\beta} + \frac{1}{4}  H\big[ Z ,Y\big] \, D^{+}_\beta Y^{+\beta}  \big) = 2 D^{+}_\beta \delta_\Lambda H\big[ Z,Y\big]  \, Y^{+\beta} + \frac{1}{2}  \delta_\Lambda H\big[ Z ,Y\big] \, D^{+}_\beta Y^{+\beta} - \nn \\-
 D^{+}_\beta H\big[ \Lambda,Y\big]  \, Y^{+\beta} - \frac{1}{4}  H\big[ \Lambda ,Y\big] \, D^{+}_\beta Y^{+\beta},
\eea
it is easier to rewrite \p{delta4} in such a way that it would involve $\delta_\Lambda Y^{+\alpha}$ only through $\delta_\Lambda H\big[Z,Y  \big]$ (with one exception) and remove it by \p{delta42} directly:
\bea\label{delta44}
-8 \delta S_{mod}&=&  \int d^6 x d^4\theta^{-} du \Big[-2\Big(  D^{+}_\beta  H\big[ Z,Y \big]  \,D^{++}\delta_\Lambda X^{+\beta} + \frac{1}{4} H\big[ Z, Y \big] \,D^{++} D^{+}_\beta\delta_\Lambda X^{+\beta}   \Big) + \nn \\
&&+2 D^{++}Z \big( D^{+}_\beta \Phi \big[ \delta_\Lambda X \big]  \,Y^{+\beta} + \frac{1}{4}\Phi \big[ \delta_\Lambda X \big]  \, D^{+}_\beta Y^{+\beta}    \big) +\nn \\
&&+ D^{++}\Lambda \big( D^{+}_\beta H\big[ Z,Y\big]  \,Y^{+\beta} + \frac{1}{4}H\big[ Z,Y\big] \, D^{+}_\beta Y^{+\beta}\Big)  -\nn \\
&&- D^{++}Z\big(  D^{+}_\beta H\big[ \Lambda,Y\big]  \,Y^{+\beta} + \frac{1}{4} H\big[ \Lambda ,Y\big] \, D^{+}_\beta Y^{+\beta}\Big) +\nn \\
&&+M\,  D^{++} D^+_\alpha \delta_\Lambda X^{+\alpha}  - D^{++}\Lambda\, M\, D^+_\gamma Y^{+\gamma} - D^{++}Z\, M\, D^+_\gamma\delta_\Lambda Y^{+\gamma} \Big].
\eea
As variation of $B_\alpha{}^\beta$ is proportional to $R_\alpha{}^\beta$ \p{trans4}, transformation of $X^{+\alpha}$ can be expected to be proportional to $Y^{+\alpha}$. Indeed, substitution $\delta_\Lambda X^{+\alpha} = \Lambda Y^{+\alpha}$ reduces \p{delta44} to
\be\label{delta45}
-8 \delta S_{mod}= \int d^6 x d^4\theta^{-} du  \Big[ M\, \Lambda\, D^{++}D^+_\gamma Y^{+\gamma} - M\, D^{++}Z \delta_\Lambda D^+_\gamma Y^{+\gamma}+ \delta_\Lambda M \, N^{+4} + M\, \delta_\Lambda N^{+4} \Big].
\ee
This expression vanishes if one takes
\be\label{deltaN4}
\delta_\Lambda N^{+4} = - \Lambda D^{++}D^+_\gamma Y^{+\gamma} + D^{++}Z\delta_\Lambda D^+_\gamma Y^{+\gamma} - \frac{1}{M}\delta_\Lambda M \, N^{+4},
\ee
and the complete transformations that extend \p{trans4} read
\bea\label{delta4ext}
\delta_\Lambda Z &=& \Lambda, \;\; \delta_\Lambda X^{+\alpha} = \Lambda \, Y^{+\alpha}, \;\; \delta_\Lambda Y^{+\alpha} =-\im  \frac{\partial^{\alpha\beta}Z}{\partial_{\mu\nu}Z \, \partial^{\mu\nu}Z}\left[ \frac{1}{2}  D^-_\beta Z\, \delta_\Lambda D^+_\gamma Y^{+\gamma} +\Lambda D^+_\beta \Phi\big[ Y\big] \right], \nn \\
\delta_\Lambda M &=& -\frac{1}{2} \left[  \Lambda \Phi\big[ Y\big]  - 2\im \Lambda \frac{D^-_\beta Z\, \partial^{\beta\gamma}Z \, D^+_\gamma \Phi\big[ Y\big]}{\partial_{\mu\nu}Z \, \partial^{\mu\nu}Z}  - \widehat{D^{--}Z }\delta_\Lambda D^+_\gamma Y^{+\gamma}\right], \nn \\
\delta_\Lambda N^{+4} &=& - \Lambda D^{++}D^+_\gamma Y^{+\gamma} + D^{++}Z \delta_\Lambda D^+_\gamma Y^{+\gamma} +\\&&+ \frac{N^{+4}}{2M} \left[  \Lambda \Phi\big[ Y\big]  - 2\im \Lambda \frac{D^-_\beta Z\, \partial^{\beta\gamma}Z \, D^+_\gamma \Phi\big[ Y\big]}{\partial_{\mu\nu}Z \, \partial^{\mu\nu}Z}  - \widehat{D^{--}Z }\delta_\Lambda D^+_\gamma Y^{+\gamma}\right].\nn
\eea
Though the $N^{+4}$ variation is singular in $M$, which does not allow to straightforwardly put transformations \p{delta4ext} on-shell, this is a price to be paid to find the transformations explicitly. If Lagrange multipliers $M$ and $N^{+4}$ were removed from the action, variations of $X^{+\alpha}$ and $Y^{+\alpha}$ would be related by equation obtained by setting $\delta_\Lambda M =0$ in \p{delta421}
\be\label{delta46}
\big( \Phi\big[ \delta_\Lambda X\big]  -  H\big[\Lambda, Y  \big] \big) - 2\im \frac{D^-_\beta Z\, \partial^{\beta\gamma}Z \, D^+_\gamma \big( \Phi\big[ \delta_\Lambda X\big]  -  H\big[\Lambda, Y  \big] \big)}{\partial_{\mu\nu}Z \, \partial^{\mu\nu}Z}  - \widehat{D^{--}Z }\delta_\Lambda D^+_\gamma Y^{+\gamma} =0.
\ee
Equation \p{delta46} does not allow, in general, to obtain $\delta_\Lambda D^+_\beta Y^{+\beta}$, as this would require division by a charged object, and should be solved with respect to $\delta_\Lambda X^{+\alpha}$, which is quite difficult. Moreover, obtained $\delta_\Lambda X^{+\alpha}$ should be substituted back to the variation of the action, and it is not possible to show just  from \p{delta46} that it vanishes. Equation \p{delta46} can be solved for $\delta_\Lambda D^+_\gamma Y^{+\gamma}$ in one particular case when the gauge parameter and variation of $X^{+\alpha}$ are chosen as $\Lambda = \widehat{D^{--}Z } \Lambda^{++}$ and $\delta X^{+\alpha} = \widehat{D^{--}Z } \Lambda^{++}Y^{+\alpha}$, so that $\widehat{D^{--}Z } $  factors out. Though such transformation leaves the action invariant, it is still not acceptable as a general solution, as such variation of $Z$ will always contain harmonics and thus would not allow to partially fix the gauge by removing harmonic dependence from components of $Z$, leaving only transformations \p{trans4}.

In the next section we partially fix gauge symmetry \p{delta4ext} by removing harmonic dependence from $Z$ to show that no new degrees of freedom appear and to calculate the component action. Alternative approach that does not rely on gauge symmetries but uses direct constraints on $Z$ is discussed in the Appendix A.

\setcounter{equation}0
\section{Equations of motion and component action}
After gauge symmetries of the supersymmetric PST action \p{modact} were established \p{deltaZW2}, \p{delta4ext}, one can find equations of motion, show that they, after partial gauge fixing, remove extra degrees of freedom contained in the harmonic expansions and finally evaluate the component action. For simplicity, let us perform latter two tasks in the bosonic limit.

Equations of motion, obtained by varying the action \p{modact}
\bea\label{modact2}
-8 S_{mod}&=&  \int d^6 x d^4\theta^{-} du \Big[ D^{+}_\beta \Phi\big[ X\big]  \,D^{++}X^{+\beta} + \frac{1}{4}\Phi\big[ X\big] \,D^{++} D^{+}_\beta X^{+\beta} -  \\
&&-2 \Big(  D^{+}_\beta H\big[ Z,Y\big]  \,D^{++}X^{+\beta} + \frac{1}{4}H\big[ Z,Y\big] \,D^{++} D^{+}_\beta X^{+\beta}   \Big) + \nn \\
&&+ D^{++}Z \Big( D^{+}_\beta H\big[ Z,Y\big]  \,Y^{+\beta} + \frac{1}{4}H\big[ Z,Y\big] \, D^{+}_\beta Y^{+\beta}\Big) +\nn \\
&&+ M \big( D^{++}D^+_\gamma X^{+\gamma} - D^{++}Z  D^+_\gamma Y^{+\gamma} + N^{+4} \big) \Big],\nn
\eea
with respect to $Z$, $M$, $N^{+4}$ and tensor multiplet prepotentials $X^{--}$, $Y^{--}$ read
\bea
\delta X^{--}:&& \;\; D^{++}\big( \Phi\big[ X \big] - H \big[ Z, Y \big] + 2 M    \big) =0, \label{Xeommod1}  \\
\delta Y^{--}:&& \;\;H\big[ Z, D^{++}X^{+\alpha} - D^{++}Z \, Y^{+\alpha}     \big] + 2D^{++}Z \,M =0,  \label{Yeommod1}  \\
\delta Z: &&\;\; D^{+}_\alpha \Phi\big[Y \big] \big( D^{++}X^{+\alpha} -D^{++}Z Y^{+\alpha} \big) +\frac{1}{4} \Phi\big[ Y\big] \big( D^{++}D^{+}_\alpha X^{+\alpha} -D^{++}Z D^{+}_\alpha Y^{+\alpha}\big) +\nn \\
&&+D^{+}_\alpha D^{++}\big( \Phi\big[ X\big] -H\big[Z,Y\big]  \big)Y^{+\alpha} + \frac{1}{4} D^{+}_\alpha Y^{+\alpha} D^{++}\big( \Phi\big[ X\big] -H\big[Z,Y\big]  \big) + \nn \\&&+ \frac{1}{2} D^{++}\big( M D^+_\beta Y^{+\beta}  \big)=0,\label{Zeommod1} \\
\delta M:&& \;\; D^{++}D^+_\gamma X^{+\gamma} - D^{++}Z  D^+_\gamma Y^{+\gamma}+ N^{+4}=0, \label{Meom1} \\
\delta N^{+4}:&&\;\; M=0.\label{Neom1}
\eea
Note that the last two equations are clearly algebraic, expressing $N^{+4}$ in terms of other superfields and setting $M$ to zero. Therefore, $M$ and $N^{+4}$ contain no new degrees of freedom. Removing $M$ from equations \p{Xeommod1}, \p{Yeommod1}, \p{Zeommod1} one obtains just the equations that follow from unmodified action \p{semifinalact}
\bea
\delta X^{--}:&& \;\; D^{++}\big( \Phi\big[ X \big] - H \big[ Z, Y \big]   \big) =0, \label{Xeommod2}  \\
\delta Y^{--}:&& \;\;H\big[ Z, D^{++}X^{+\alpha} - D^{++}Z \, Y^{+\alpha}     \big]  =0,  \label{Yeommod2}  \\
\delta Z: &&\;\; D^{+}_\alpha \Phi\big[Y \big] \big( D^{++}X^{+\alpha} -D^{++}Z Y^{+\alpha} \big) +\frac{1}{4} \Phi\big[ Y\big] \big( D^{++}D^{+}_\alpha X^{+\alpha} -D^{++}Z D^{+}_\alpha Y^{+\alpha}\big) + \nn \\
&&+D^{+}_\alpha D^{++}\big( \Phi\big[ X\big] -H\big[Z,Y\big]  \big)Y^{+\alpha} + \frac{1}{4} D^{+}_\alpha Y^{+\alpha} D^{++}\big( \Phi\big[ X\big] -H\big[Z,Y\big]  \big)=0.\label{Zeommod2}
\eea
To show that no new degrees of freedom appear and to calculate the component action one needs to analyze just the first two equations \p{Xeommod2}, \p{Yeommod2}.
Let us perform this analysis in the bosonic limit by substituting appropriate $\theta$-expansions of superfields $X^{+\alpha}$, $Y^{+\alpha}$ and $Z$. Note that invariance of the action with respect to gauge transformations \p{delta4ext} allows to take the $Z$ superfield in its short form, partially fixing the gauge and leaving only transformations of type \p{trans4}:
\bea\label{XYthetaexp}
X^{+\alpha} &\approx& \theta^{\beta} \big( \delta_\beta^\alpha q + B_\alpha{}^\beta  \big) + \big( \theta^{+3} \big)_\beta E^{-2 \beta\alpha} + \theta^{-\alpha} \big( f^{+2} + \theta^{\mu}\theta^\nu a_{\mu\nu} + \big(\theta^{+4}\big) C^{-2}   \big), \nn \\
Y^{+\alpha} &\approx& \theta^{\beta} \big( \delta_\beta^\alpha c + R_\alpha{}^\beta  \big) + \big( \theta^{+3} \big)_\beta K^{-2 \beta\alpha} + \theta^{-\alpha} \big( g^{+2} + \theta^{\mu}\theta^\nu b_{\mu\nu} + \big(\theta^{+4}\big) D^{-2}   \big), \nn \\
Z &\approx& z, \;\; \partial^{++}z=0.
\eea

Expanded in components, \p{Xeommod2} produces five equations:
\bea
&&\partial^{++} \left( q + \frac{1}{2}\partial^{--}f^{+2}   \right) =0, \label{Xeommod31}  \\
&&8 \im \partial_{\mu\nu} \left( q + \frac{1}{2}\partial^{--}f^{+2}   \right) + \partial^{++} \left( 4 \partial^{--}a_{\mu\nu} + 2 E^{-2}_{[\mu\nu]}    \right) +\nn \\&&+ 4 a_{\mu\nu} -4 \im \partial_{[\mu\sigma}B_{\nu]}{}^\sigma - 4\im \partial_{\mu\nu}q+ 4\im \partial_{[\mu\sigma}z R_{\nu]}{}^\sigma + 4\im \partial_{\mu\nu}z c=0, \label{Xeommod32}\\
&&4\partial^{++}   \partial^{--}C^{-2}  -2\im \partial^{\mu\nu} \left( 4 \partial^{--} a_{\mu\nu} + 2 E^{-2}_{\mu\nu}  \right) - \left( -8C^{-2} - 4\im \partial_{\mu\nu}E^{-2\mu\nu} + 4\im\partial_{\mu\nu}z K^{-2\mu\nu}  \right)=0, \label{Xeommod33}  \\
&&\partial^{++}\left( 4 a_{\mu\nu} -4\im \partial_{\mu\sigma} B_{\nu}{}^\sigma -4\im \partial_{\mu\nu}q + 4\im \partial_{\mu\alpha}z R_{\nu}{}^\alpha + 4\im \partial_{\mu\nu} z\, c    \right)=0, \label{Xeommod34} \\
&&-2\im \partial^{\nu\lambda}\left( 4 a_{\mu\nu} -4 \im \partial_{\mu\nu} q -4\im \partial_{\mu\sigma} B_{\nu}{}^\sigma  + 4\im \partial_{\mu\alpha}z R_{\nu}{}^\alpha + 4\im \partial_{\mu\nu}z \, c  \right)+ \nn \\&&+\partial^{++}\left( -2 \delta_\mu^\lambda    C^{-2} + 4\im \partial_{\alpha\mu}E^{-2\lambda\alpha}  -4\im \partial_{\alpha\mu}z \, K^{-2\lambda\alpha} \right)=0. \label{Xeommod35}
\eea
Equation \p{Yeommod2} should be expanded in components, too. Note that applying $D^+_\alpha$ to \p{Yeommod2}  one can obtain
\bea\label{Yeommod3}
&&D^{++}X^{+\alpha} - D^{++}Z Y^{+\alpha} =- \frac{\im}{2} \frac{\partial^{\alpha\gamma}Z D^-_\gamma Z}{\partial_{\mu\nu}Z \partial^{\mu\nu}Z} \left( D^{++}D^+_\beta X^{+\beta} - D^{++}Z D^+_\beta Y^{+\beta} \right)\;\; \Rightarrow \nn \\
&& \widehat{D^{--}Z} \big( D^{++}D^+_\beta X^{+\beta} - D^{++}Z D^+_\beta Y^{+\beta}  \big)=0.
\eea
In our gauge \p{XYthetaexp} $\widehat{D^{--}Z}$ \p{delta421} equals zero. Therefore, \p{Yeommod2} contains only two independent component equations that come from its nonanalytic part:
\bea
&&  -4\im \partial_{\mu\alpha}z \big( \partial^{++}B_\nu{}^\alpha + \partial^{++}q\delta_\nu^\alpha + f^{+2}\delta_\nu^\alpha \big)=0,\label{Yeommod31}  \\
&& 4\im \partial_{\alpha\mu}z  \left( \partial^{++}E^{-2\lambda\alpha} +2\im \partial^{\lambda\sigma}B_\sigma{}^\alpha +2\im \partial^{\lambda\alpha} q +2 a^{\lambda\alpha} -2\im \partial^{\lambda\sigma}z \big( R_\sigma{}^\alpha + \delta_\sigma^\alpha c \big) \right)=0. \label{Yeommod32}
\eea
Note that one can multiply each of these equations by $\partial^{\rho\mu}z$ and divide by appearing $\partial_{\alpha\beta}z\, \partial^{\alpha\beta}z \neq 0$. Thus factors of $\partial_{\mu\alpha}z$ can be forgotten in both equations, and one should put to zero contents of brackets in \p{Yeommod31}, \p{Yeommod32}. Trace and traceless parts of bracket in \p{Yeommod31} are equal to zero independently:
\be\label{Yeommod311}
\partial^{++}B_\alpha{}^\beta =0, \;\; \partial^{++}q + f^{+2} =0.
\ee
Therefore, $B_\alpha{}^\beta$ does not contain any extra degrees of freedom, related to the harmonics. Combining \p{Yeommod311} with \p{Xeommod31}, one can exclude $\partial^{++}q$ and obtain
\be\label{feq}
-f^{+2} +\frac{1}{2}\partial^{++}\partial^{--}f^{+2}=0 \;\; \Rightarrow \;\; \partial^{--}\partial^{++}f^{+2}=0 \;\; \Rightarrow f^{+2} = f^{ij}u^+_i u^+_j.
\ee
Equations \p{Xeommod31}, \p{Yeommod31} are certainly algebraic and do not restrict dynamics of the main fields $B_\alpha{}^\beta$ and $q$.

Equations \p{Xeommod32}, \p{Xeommod34} are algebraic also. Part of \p{Xeommod34}, symmetric in $\mu$, $\nu$ together with \p{Yeommod311} implies restriction on harmonic-dependent part of $R_\alpha{}^\beta$:
\be\label{Xeommod341}
\partial_{(\mu\sigma}z \partial^{++}R_{\nu)}{}^\sigma =0 \;\; \Rightarrow \;\;R_\alpha{}^\beta = \big(R_0\big)_\alpha{}^\beta + \partial_{\alpha\gamma}z S^{[\beta\gamma]} - \frac{1}{4}\delta_\alpha^\beta \partial_{\mu\nu} z S^{[\mu\nu]},
\ee
where $R_0$ does not depend on harmonics and $S^{\mu\nu}$ is proportional to harmonics.

The antisymmetric part of \p{Xeommod34} implies that \p{Xeommod32} can be split into two parts that vanish independently. First one is proportional to harmonics
\be\label{Xeommod322}
\partial^{++} \left( 4 \partial^{--}a_{\mu\nu} + 2 E^{-2}_{[\mu\nu]}    \right) =0 \;\; \Rightarrow \;\;  E^{-2}_{[\mu\nu]} =-2\partial^{--}a_{\mu\nu},
\ee
while other does not contain harmonics at all:
\bea\label{Xeommod321}
8 \im \partial_{\mu\nu} \left( q + \frac{1}{2}\partial^{--}f^{+2}   \right) +  4 a_{\mu\nu} -4 \im \partial_{[\mu\sigma}B_{\nu]}{}^\sigma - 4\im \partial_{\mu\nu}q+ 4\im \partial_{[\mu\sigma}z R_{\nu]}{}^\sigma + 4\im \partial_{\mu\nu}z c=0\;\; \Rightarrow \nn \\
a_{\mu\nu} = \im \partial_{[\mu\sigma}B_{\nu]}{}^\sigma + \im \partial_{\mu\nu}q -2\im \partial_{\mu\nu} \left( q + \frac{1}{2}\partial^{--}f^{+2}   \right) - \im \partial_{[\mu\sigma}z R_{\nu]}{}^\sigma - \im \partial_{\mu\nu}z c.
\eea
Using \p{Xeommod321} to simplify equation \p{Xeommod35}, one extracts harmonic-independent portion, which naturally splits into trace and traceless parts
\be\label{Xeommod352}
\partial^{\nu\lambda} \big( \partial_{(\mu\sigma} B_{\nu)}{}^\sigma  - \partial_{(\mu\alpha}z R_{\nu)}{}^\alpha  \big)=0, \;\;  \partial_{\alpha\beta}\partial^{\alpha\beta}\left( q + \frac{1}{2}\partial^{--}f^{+2}   \right)=0.
\ee
Both of them are true physical equations of motion. The first one in \p{Xeommod352} after exclusion of auxiliary field $R_\alpha{}^\beta$ leads to self-duality equation in the same way as the equation of motion that comes from the PST action. We do not either use or analyze \p{Xeommod352} further. Remaining part of \p{Xeommod35} leads to algebraic constraint. Together with \p{Xeommod33} it could be solved for $C^{-2}$ and $K^{-2\alpha\beta}$, though explicit result is unimportant.

The only remaining equation is \p{Yeommod32}. Simplified with help of other algebraic equations, it can be split into symmetric and antisymmetric parts:
\bea
\partial^{++}E^{-2(\lambda\beta)} + 2\im \partial^{(\lambda\sigma}B_\sigma{}^{\beta)} -2\im \partial^{(\lambda\sigma}z R_\sigma{}^{\beta)}=0, \label{Yeommod321} \\
-2 \partial^{\lambda\beta}z \big( \partial^{++}\partial^{--}c -2c \big) -2 \partial^{[\lambda\sigma}z \, \partial^{++}\partial^{--}R_\sigma{}^{\beta]} =0.\label{Yeommod322}
\eea
Note that $E^{-2(\lambda\beta)}=0$ as the rest of equation \p{Yeommod321} does not depend on harmonics. Therefore, \p{Yeommod321} reduces to an algebraic equation that should be solved for $R_\alpha{}^\beta$. As we want to keep $R_\alpha{}^\beta$ field in the action, we do not solve  \p{Yeommod321} explicitly. Last equation \p{Yeommod322}, being multiplied by $\partial_{\lambda\beta}z$, reduces to $\partial^{++}\partial^{--}c -2c =0$, resulting in $c = c^{ij}u^+_i \, u^-_j$. The remaining term in \p{Yeommod322} implies $\partial^{++}\partial^{--}S^{\lambda\beta}=0$, or simply $S^{\lambda\beta}=0$, as it was assumed to be proportional to harmonics \p{Xeommod341}.

We see that equations \p{Xeommod2}, \p{Yeommod2} remove most of harmonic dependence of the components of the superfields $X^{+\alpha}$, $Y^{+\alpha}$, which makes possible calculation of the component action. Performing integration in \p{semifinalact} over $\theta$ variables,
\bea\label{actintegr1}
-8S &=& \int d^6 x du \left[ 8 \left( q + \frac{1}{2}\partial^{--}f^{+2}   \right) \left(\partial^{++}C^{-2} - 2\im \partial_{\mu\nu}a^{\mu\nu} \right) - \right.\nn \\ &&\left. -2 \left( \partial^{++}a^{\mu\nu}+ \im \partial^{\mu\nu}f^{+2}  \right)\left( 4 \partial^{--}a_{\mu\nu} + 2 E^{-2}_{\mu\nu}  \right)- \nn \right. \\ &&\left. -\left( 4 a_{\alpha\mu} - 4\im \partial_{\alpha\rho}B_\mu{}^\rho -4\im \partial_{\alpha\mu}q  + 8\im \partial_{\alpha\rho} z R_\mu{}^\rho + 8\im \partial_{\alpha\mu}z \, c    \right) \times \right. \nn \\ &&\left. \times \left( \partial^{++}E^{-2\mu\alpha} +2\im \partial^{\mu \sigma}B_\sigma{}^\alpha + 2\im \partial^{\mu\alpha}q +2 a^{\mu\alpha}   \right)+ \right. \nn \\ &&\left. +8 \partial^{\mu\nu}z \partial_{\rho\sigma}z \big( R_\mu{}^\rho + \delta_\mu^\rho c \big)\big( R_\nu{}^\sigma + \delta_\nu^\sigma c \big) \right].
\eea
After applying algebraic equations, only harmonic-independent fields remain, making harmonic integration trivial:
\bea\label{actintegrfin}
S = \int d^6 x \left[ -2 \partial_{\mu\nu}q_0 \partial^{\mu\nu}q_0 + \partial^{(\mu\sigma}B_\sigma{}^{\nu)} \partial_{(\mu\rho}B_{\nu)}{}^\rho -2 \partial^{(\mu\sigma}B_\sigma{}^{\nu)} \partial_{(\mu\rho}z R_{\nu)}{}^\rho + \partial_{(\mu\rho}z R_{\nu)}{}^\rho \, \partial^{(\mu\sigma}z R_\sigma{}^{\nu)} \right].
\eea
Here $q_0 = q + 1/2 \partial^{--}f^{+2}$. As expected, the obtained action coincides, up to term with $q_0$, with the polynomial PST action \p{PSTMact2}.

\setcounter{equation}0
\section{Conclusion}
In the present paper we constructed the superfield action for the free $N=(1,0)$, $d=6$ tensor multiplet, which generalizes the Pasti-Sorokin-Tonin action for self-dual tensor field. As standard description of the $d=6$ tensor multiplet involves on-shell superfields, we employed superfields defined on $N=(1,0)$, $d=6$ harmonic superspace, with the action given by the integral over analytic subspace. Our construction was inspired by the polynomial form of the PST action found by Mkrtchyan \cite{mkrtchyan}. In this formulation, one can provide a superfield analog to each of three terms of the bosonic action. The superfields involved in the construction are the spinor potentials $X^{+\alpha}$, $Y^{+\alpha}$, associated with physical and auxiliary tensor multiplets, analytic superfield $Z$, first component of which is the gauge PST scalar, and auxiliary analytic superfields $M$ and $N^{+4}$. We provided superfield generalizations of all gauge symmetries of the PST action. After partial gauge fixing, we checked in the bosonic limit that algebraic equations of motion that follow from our action remove all the auxiliary fields contained in the harmonic expansions of tensor superfields and gauge analytic ones, while the remaining equations are dynamical and lead to self-duality constraint on the 2-form field in the same manner as in the original bosonic case. Therefore, our action contains correct number of degrees of freedom and is the proper one for the tensor multiplet.

The present off-shell construction could be used as a framework to study couplings of the tensor multiplet to matter and to itself, and, most importantly, nonabelian generalization of the tensor multiplet. It would be also interesting to construct analogous mechanism for $N=(1,0)$, $d=6$ supergravity, which involves the tensor field of opposite duality.

Let us finally note that once analytic superfield $Z$ is introduced one can use it to split the nonanalytic potential $X^{+\alpha}$, $D^{+}_\beta X^{+\alpha}\sim \delta_\beta^\alpha$ into two unconstrained analytic ones, ${\widetilde {X}}^{+\alpha}$ and $X^{++}$. This points to a possibility of describing the tensor multiplet in terms of purely analytic unconstrained superfields following ideas of Buchbinder, Ivanov and Zaigraev who found that not only Yang-Mills theory, hypermultiplet and supergravity but also higher spin theories are naturally described by analytic superfields \cite{BIZ}.

\section*{Acknowledgments}
The work was supported by Russian Foundation for Basic Research, grant No 20-52-12003.

\setcounter{equation}0
\def\theequation{A.\arabic{equation}}
\section*{Appendix A. Alternative constraints}
As an alternative to fixing one of the gauge symmetries of the action \p{modact}, one may consider adding constraints that remove excessive components from superfield $Z$. Such constraints are not easy to find, as they should not affect the harmonic-independent part of the first component of $Z$, otherwise important transformations \p{trans4} could not be extended to the supersymmetric case. We propose the following constraints
\be\label{Zconstr}
\big( D^{++}  \big)^3 Z = 0, \;\; {\widehat {D^{--}Z}} = D^{--}Z +\im \frac{D^-_\rho Z \, D^-_\sigma Z \, \partial^{\rho\sigma}Z}{\partial_{\mu\nu}Z \partial^{\mu\nu}Z} =0.
\ee
They can be added to the action \p{semifinalact} with analytic Lagrange multipliers:
\be\label{constrnew}
S_{new} = S + \int d^6 x d^4 \theta^{-} du \Big[ \Omega^{-2}_1 \big( D^{++}  \big)^3 Z + \Omega^{+6}_2 {\widehat {D^{--}Z}}   \Big].
\ee
As added terms \p{constrnew} do not contain either $X^{+\alpha}$ or $Y^{+\alpha}$, equations obtained by varying $S$ \p{semifinalact} with respect to $X^{--}$ and $Y^{--}$ are unmodified \p{Xeommod2} and \p{Yeommod2}. Varying with respect to $\Omega_1$, $\Omega_2$, one recovers \p{Zconstr}. Finally, varying with respect to $Z$ one finds \p{Zeommod2} plus terms with Lagrange multipliers. We would not use it explicitly anyway, as \p{Xeommod2}, \p{Yeommod2} and \p{Zconstr} would be sufficient for our purposes.

Just as we did previously, let us analyze equations \p{Xeommod2}, \p{Yeommod2}, \p{Zconstr} in the bosonic limit, neglecting fermions but keeping both charged and uncharged bosonic components and not restricting their harmonic dependence. Thus we consider the $\theta$-expansions of $X^{+\alpha}$, $Y^{+\alpha}$ as defined in \p{XYthetaexp} while the $Z$ superfield reads
\bea\label{Zthetaexp}
Z&\approx& z + \theta^\mu \theta^\nu d^{-2}_{\mu\nu} +\big(\theta^{+4}\big) L^{-4}.
\eea

\subsection{$Z$ equation}
At first, let us show that \p{Zconstr} remove unnecessary components from $Z$ superfield. The first constraint $\big( D^{++}  \big)^3Z=0 $ is linear. After applying three derivatives to $Z$ \p{Zthetaexp}, one obtains
\bea\label{constr1}
\big( \partial^{++} \big)^3 z =0, \;\; 3\im \big( \partial^{++} \big)^2 \partial_{\mu\nu}z + \big( \partial^{++} \big)^3 d^{-2}_{\mu\nu} =0, \nn\\
 \big( \partial^{++} \big)^3 L^{-4} + 4\partial^{++} \partial_{\mu\nu}\partial^{\mu\nu} z - 6\im \big( \partial^{++} \big)^2 \partial_{\mu\nu}d^{-2\mu\nu} =0.
\eea
From the first of these equations it follows that $z$ has the harmonic expansion
\be\label{constr11}
z = z_0 + u^+_i u^-_j z^{(ij)} + u^+_i u^+_j u^-_k u^-_l z^{(ijkl)}.
\ee
From the second one, it follows that
\be\label{constr12}
d^{-2}_{\mu\nu} = d^{(ij)\mu\nu}u^-_i u^-_j - \im \partial_{\mu\nu}z^{(ijkl)} u^+_i u^-_j u^-_k u^-_l.
\ee
The third equation reduces to two projections, proportional to $u^2$ and $u^4$, which result in
\be\label{constr13}
L^{-4} = 24 \partial_{\mu\nu}\partial^{\mu\nu} z^{(ijkl)}u^-_i u^-_j u^-_k u^-_l, \;\; 4 \partial_{\mu\nu}\partial^{\mu\nu} z^{(ij)} -12\im \partial^{\mu\nu}d_{\mu\nu}^{(ij)} =0.
\ee
Therefore, $d^{(ij)}_{\mu\nu}$ can be taken as
\be\label{constr14}
d^{(ij)}_{\mu\nu} = -\frac{\im}{3} \partial_{\mu\nu}z^{(ij)} + {\tilde d}^{(ij)}_{\mu\nu}, \;\; \partial^{\mu\nu} {\tilde d}^{(ij)}_{\mu\nu} =0.
\ee
Thus we conclude that constraint $\big( D^{++}  \big)^3Z=0 $ is off-shell and imposes no equations on the field $z_0$, relating other fields to harmonic expansion of $z$ (with exception of ${\tilde d}^{(ij)}_{\mu\nu}$).

The second constraint $ {\widehat {D^{--}Z}} =0$, in spite of being nonlinear, drastically simplifies superfield $Z$. It reduces to three bosonic equations\footnote{In general case, $\partial^{--}z$ will be proportional to the fermions},
\bea\label{constr2}
\partial^{--}z =0, \;\; \partial^{--}d^{-2}_{\mu\nu} + 4\im \frac{\partial^{\alpha\beta} z d^{-2}_{\alpha\mu} d^{-2}_{\beta\nu}}{\partial_{\rho\sigma}z\partial^{\rho\sigma} z} =0, \nn \\
\partial^{--}L^{-4} + 4\im \frac{\partial^{\alpha\beta}z\, d^{-2}_{\alpha\beta} L^{-4}}{\partial_{\rho\sigma}z\partial^{\rho\sigma} z} - 8\im \frac{\partial^{\alpha\beta}d^{-2\mu\nu} d^{-2}_{\alpha\mu}d^{-2}_{\beta\nu}}{\partial_{\rho\sigma}z\partial^{\rho\sigma} z} + 16\im \frac{\partial^{\alpha\beta} z\, \partial^{\lambda\tau} z\, \partial_{\lambda\tau}d^{-2\mu\nu}d^{-2}_{\alpha\mu}d^{-2}_{\beta\nu}}{\big(\partial_{\rho\sigma}z\partial^{\rho\sigma} z   \big)^2}=0.
\eea
The first of these equations implies that $z$ is harmonic-independent, which together with \p{constr12}, \p{constr13}, \p{constr14} puts to zero $L^{-4}$ and reduces $d^{-2\mu\nu}$ to ${\tilde d}^{-2\mu\nu}$. Therefore, $\partial^{--}d^{-2}_{\mu\nu}$ disappears from the second equation, and it becomes an algebraic constraint
\be\label{constr21}
\partial^{\alpha\beta} z \,d^{-2}_{\alpha\mu} d^{-2}_{\beta\nu} =0 \;\; \Rightarrow \;\; \frac{1}{2} \big( \partial_{\rho\sigma}z d^{-2\rho\sigma}  \big)d^{-2}_{\mu\nu} - \frac{1}{4}\partial_{\mu\nu}z \big( d^{-2}_{\rho\sigma}d^{-2\rho\sigma} \big) =0.
\ee
Multiplying this equation by $\partial^{\mu\nu} z$ or by $d^{-2\mu\nu}$, one can conclude that both $\partial_{\rho\sigma}z d^{-2\rho\sigma}$ and $d^{-2}_{\rho\sigma}d^{-2\rho\sigma} $ are equal to zero, which is sufficient to satisfy this equation. Finally, the third equation in \p{constr2} is satisfied identically if all others are taken into account. The doubly constrained superfield $Z$, therefore, reads
\be\label{Zsimpl}
Z \approx z + \theta^{+\mu}\theta^{+\nu} d^{-2}_{\mu\nu}, \;\; \partial^{--}z =0, \;\; \partial^{--}d^{-2\mu\nu} =0, \;\; d^{-2}_{\rho\sigma}d^{-2\rho\sigma} =0, \;\; \partial_{\rho\sigma}z d^{-2\rho\sigma} =0, \;\; \partial_{\rho\sigma} d^{-2\rho\sigma}=0.
\ee
As constraints \p{Zconstr} result in algebraic equations, the Lagrange multipliers $\Omega_1$ and $\Omega_2$ do not give rise to independent degrees of freedom.

\subsection{$X$ equation}
Component expansions of $X$ and $Y$ equations \p{Xeommod2}, \p{Yeommod2} are modified compared to \p{Xeommod31}-\p{Xeommod35}, \p{Yeommod31}, \p{Yeommod32} by the presence of $d^{-2}_{\alpha\beta}$. $X$ equation, expanded in components, now reads
\bea
&&\partial^{++} \left( q + \frac{1}{2}\partial^{--}f^{+2}   \right) =0, \label{Xeq1}  \\
&&8 \im \partial_{\mu\nu} \left( q + \frac{1}{2}\partial^{--}f^{+2}   \right) + \partial^{++} \left( 4 \partial^{--}a_{\mu\nu} + 2 E^{-2}_{[\mu\nu]} + 4 d^{-2}_{[\mu\alpha} R_{\nu]}{}^\alpha + 4 d^{-2}_{\mu\nu}c   \right) +\nn \\&&+ 4 a_{\mu\nu} -4 \partial_{[\mu\sigma}B_{\nu]}{}^\sigma - 4 g^{+2} d^{-2}_{\mu\nu} + 4\im \partial_{[\mu\sigma}z R_{\nu]}{}^\sigma + 4\im \partial_{\mu\nu}z c=0, \label{Xeq2}\\
&&\partial^{++} \left( 4 \partial^{--}C^{-2} -4 d^{-2}_{\mu\nu} K^{-2\mu\nu}  \right)-2\im \partial^{\mu\nu} \left( 4 \partial^{--} a_{\mu\nu} + 2 E^{-2}_{\mu\nu} +4 d^{-2}_{[\mu\alpha}R_{\nu]}{}^\alpha + 4 d^{-2}_{\mu\nu} c \right) - \nn \\&&- \left( -8C^{-2} - 4\im \partial_{\mu\nu}E^{-2\mu\nu} - 8 d^{-2}_{\mu\nu} b^{\mu\nu} + 4\im\partial_{\mu\nu}z K^{-2\mu\nu} + 8\im \partial_{\alpha\mu}d^{-2\beta\mu} R_\beta{}^\alpha \right)=0, \label{Xeq3}  \\
&&\partial^{++}\left( 4 a_{\mu\nu} -4\im \partial_{\mu\sigma} B_{\nu}{}^\sigma -4\im \partial_{\mu\nu}q -4 g^{+2} d^{-2}_{\mu\nu} + 4\im \partial_{\mu\alpha}z R_{\nu}{}^\alpha + 4\im \partial_{\mu\nu} z\, c    \right)=0, \label{Xeq4} \\
&&-2\im \partial^{\nu\lambda}\left( 4 a_{\mu\nu} -4 \im \partial_{\mu\nu} q -4\im \partial_{\mu\sigma} B_{\nu}{}^\sigma -4 g^{+2}d^{-2}_{\mu\nu} + 4\im \partial_{\mu\alpha}z R_{\nu}{}^\alpha + 4\im \partial_{\mu\nu}z \, c  \right)+ \nn \\&&+\partial^{++}\left( -2 \delta_\mu^\lambda    C^{-2} + 4\im \partial_{\alpha\mu}E^{-2\lambda\alpha} + 8 d^{-2}_{\mu\nu}b^{\nu\lambda} -4\im \partial_{\alpha\mu}z \, K^{-2\lambda\alpha} + 8\im \partial_{\alpha\mu}d^{-2\beta\lambda} R_\beta{}^\alpha\right)=0. \label{Xeq6}
\eea
The analysis of these equations is mostly similar compared to one of gauge-fixed equations. The first of these equations \p{Xeq1} implies that the multiplet contains harmonic-independent quantity
\be\label{tildeq}
q_0 = q + \frac{1}{2}\partial^{--}f^{+2}, \;\; \partial^{++}q_0 =0.
\ee
Looking at equation \p{Xeq2}, one can notice with help of \p{Xeq1} and \p{Xeq4} that its first bracket and second line do not depend on harmonics while the rest is proportional to harmonics. These parts should vanish independently:
\bea
a_{\mu\nu} = \im \partial_{[\mu\sigma} B_{\nu]}{}^\sigma +\im \partial_{\mu\nu} q - \im\partial_{[\mu\alpha}z R_{\nu]}{}^\alpha -\im \partial_{\mu\nu} z\, c -2\im \partial_{\mu\nu}q_0\;\; \label{aexpr}, \\
E^{-2}_{[\mu\nu]} = -2 \partial^{--}a_{\mu\nu} -2 d^{-2}_{[\mu\alpha} R_{\nu]}{}^\alpha -2d^{-2}_{\mu\nu}\, c=0. \label{Eexpr}
\eea
Third equation \p{Xeq3}, just as the symmetric part of \p{Xeq4}, is an algebraic one and can be solved for $C^{-2}$. Finally, \p{Xeq6}, after taking \p{aexpr}, \p{Eexpr}, \p{Xeq4} into account, splits into harmonic-dependent algebraic equation, solvable for $K^{-2\alpha\beta}$, and harmonic-independent part, which is physical equation of motion. Consequences of  \p{Xeq6} are not useful in calculation of the component action anyway.

\subsection{$Y$ equation}
Remaining important equation is the $Y^{+\alpha}$ one \p{Yeommod2}. Its analytic part vanishes due to constraints on $Z$ \p{Yeommod3}, \p{Zconstr}. The first relevant component equation reads
\be\label{Yeom4}
4 d^{-2}_{\mu\nu} \partial^{++}f^{+2} -4\im \partial_{\mu\alpha}z \big( \partial^{++}B_\nu{}^\alpha + \partial^{++}q\delta_\nu^\alpha + f^{+2}\delta_\nu^\alpha \big)=0.
\ee
Multiplying it by $\partial^{\mu\nu}z$, due to properties of $d^{-2}_{\mu\nu}$ one finds $f^{+2}+\partial^{++}q =0$. Comparing with \p{Xeq1}, one obtains, as before,
\be\label{feq2}
 f^{+2} = f^{ij}u^+_i u^+_j.
\ee
In particular, $\partial^{++}f^{+2}=0$ and equation \p{Yeom4} reduces to just
\be\label{Yeom5}
\partial_{\mu\alpha}z \, \partial^{++}B_\nu{}^\alpha =0 \;\; \Rightarrow \;\; \partial^{++}B_\beta{}^\alpha =0.
\ee
Therefore, $B_\beta{}^\alpha$ does not depend on harmonics, and additional degrees of freedom do not appear. Symmetric part of equation \p{Xeq4} then implies
\be\label{Yeq51}
\partial_{(\mu\sigma}z \partial^{++}R_{\nu)}{}^\sigma =0 \;\; \Rightarrow \;\;R_\alpha{}^\beta = \big(R_0\big)_\alpha{}^\beta + \partial_{\alpha\gamma}z S^{[\beta\gamma]} - \frac{1}{4}\delta_\alpha^\beta \partial_{\mu\nu} z S^{[\mu\nu]},
\ee
where $R_0$ does not depend on harmonics and $S^{\mu\nu}$ is proportional to harmonics.
Final equation also gets modified compared to \p{Yeommod32}:
\bea\label{Yeq52}
-8 d^{-2}_{\mu\nu} \left( \partial^{++}a^{\nu\lambda} + \im \partial^{\nu\lambda}f^{+2} -\im g^{+2}\partial^{\nu\lambda}z -g^{+2}\partial^{++}d^{-2\nu\lambda}    \right) +\nn \\+ 4\im \partial_{\alpha\mu}z  \left( \partial^{++}E^{-2\lambda\alpha} +2\im \partial^{\lambda\sigma}B_\sigma{}^\alpha +2\im \partial^{\lambda\alpha} q +2 a^{\lambda\alpha} -2 \big(\im \partial^{\lambda\sigma}z +\partial^{++}d^{-2\lambda\sigma} \big)\big( R_\sigma{}^\alpha + \delta_\sigma^\alpha c \big) \right)=0
\eea
Multiplying it by $\partial^{\alpha\beta}z$, one can split this equation into antisymmetric and symmetric parts:
\bea\label{Yeq53}
-2 \partial^{\lambda\beta}z \big( \partial^{++}\partial^{--}c -2c \big) -4\im \partial^{++}\big( c\, d^{-2\lambda\beta}  \big) -2\im \partial^{++}\partial^{--} \big( g^{+2}d^{-2\lambda\beta}  \big)+\nn \\+4\im d^{-2[\lambda\sigma} \partial^{++}R_\sigma{}^{\beta]}
-2 \partial^{[\lambda\sigma}z \, \partial^{++}\partial^{--}R_\sigma{}^{\beta]} =0, \nn \\
E^{-2(\lambda\beta)} = d^{-2(\lambda\sigma}R_\sigma{}^{\beta)}.
\eea
These equations should be used in calculation of the component action. Performing integration in \p{semifinalact} over $\theta$ variables,
\bea\label{actintegr11}
-8S &=& \int d^6 x du \left[ 8 q_0 \partial^{++}C^{-2} - 16\im q_0 \partial_{\mu\nu}a^{\mu\nu}-\right. \nn \\ &&\left. -2 \left( \partial^{++}a^{\mu\nu}+ \im \partial^{\mu\nu}f^{+2}  \right)\left( 4 \partial^{--}a_{\mu\nu} + 2 E^{-2}_{\mu\nu} + 8 \big( R_\mu{}^\alpha + \delta_\mu^\alpha c \big) d^{-2}_{\alpha\nu}  \right)- \nn \right. \\ &&\left. -\left( 4 a_{\alpha\mu} - 4\im \partial_{\alpha\rho}B_\mu{}^\rho -4\im \partial_{\alpha\mu}q - 8g^{+2}d^{-2}_{\alpha\mu} + 8\im \partial_{\alpha\rho} z R_\mu{}^\rho + 8\im \partial_{\alpha\mu}z \, c    \right) \times \right. \nn \\&& \left. \times \left( \partial^{++}E^{-2\mu\alpha} +2\im \partial^{\mu \sigma}B_\sigma{}^\alpha + 2\im \partial^{\mu\alpha}q +2 a^{\mu\alpha}   \right)+ \right. \nn \\ &&\left. +16 \left( \im\partial^{\mu\nu}z + \partial^{++}d^{-2\mu\nu}  \right) g^{+2} d^{-2}_{\mu\beta}\big( R_\nu{}^\beta + \delta_\nu^\beta c\big) \right. +\nn \\ &&\left.+8 \big( \partial^{\mu\nu}z -\im \partial^{++}d^{-2\mu\nu}  \big)\partial_{\rho\sigma}z \big( R_\mu{}^\rho + \delta_\mu^\rho c \big)\big( R_\nu{}^\sigma + \delta_\nu^\sigma c \big) \right].
\eea
After taking the algebraic equations of motion into account, one can reduce \p{actintegr1} to
\bea\label{actintegr2}
-8S = \int d^6 x du \left[ 16 \partial_{\mu\nu}q_0 \partial^{\mu\nu}q_0 - 8 \partial^{(\mu\sigma}B_\sigma{}^{\nu)} \partial_{(\mu\rho}B_{\nu)}{}^\rho +16 \partial^{(\mu\sigma}B_\sigma{}^{\nu)} \partial_{(\mu\rho}z R_{\nu)}{}^\rho -\right.  \nn \\ \left.- 8 \partial_{(\mu\rho}z R_{\nu)}{}^\rho \, \partial^{(\mu\sigma}z R_\sigma{}^{\nu)} + 8\im \partial_{\mu\rho }z R_\nu{}^\rho R_\sigma{}^\nu \partial^{++}d^{-2\mu\sigma} \right].
\eea
As $R_\alpha{}^\beta$ has structure \p{Yeq51}, one can check that $S$-terms in $R_\alpha{}^\beta$ cancel, and whole last term can be presented as a total harmonic derivative
\be\label{lastterm}
\partial^{++}\big( \partial_{\mu\rho}z \big(R_0  \big)_\nu{}^\rho \big(R_0  \big)_\sigma{}^\nu \, d^{-2\mu\sigma} \big),
\ee
and, therefore, makes no contribution to the integral. Also $S$ cancels from all other terms in the action, making harmonic integration trivial. Therefore, we obtain the action in its expected form
\be\label{actintegrfin2}
S = \int d^6 x \left[ -2 \partial_{\mu\nu}q_0 \partial^{\mu\nu}q_0 + \partial^{(\mu\sigma}B_\sigma{}^{\nu)} \partial_{(\mu\rho}B_{\nu)}{}^\rho -2 \partial^{(\mu\sigma}B_\sigma{}^{\nu)} \partial_{(\mu\rho}z \big(R_0\big)_{\nu)}{}^\rho + \partial_{(\mu\rho}z \big(R_0\big)_{\nu)}{}^\rho \, \partial^{(\mu\sigma}z \big(R_0\big)_\sigma{}^{\nu)} \right].
\ee
Thus we see that it is possible to show that the action \p{semifinalact} leads to, after minor modification, to the polynomial PST action even without invoking gauge symmetries. It is also possible that used constraints are stronger than minimal needed, and just ${\widehat{D^{--}Z}}=0$ or analogous condition would suffice. This is a question for further study. It is also desirable to find gauge transformations of the action with constraints \p{constrnew}, which could be nonsingular if appropriate restrictions on $\delta Z$ were enforced.

\end{document}